# Sleep Stage Classification: Scalability Evaluations of Distributed Approaches

Şerife Açıkalın, Süleyman Eken, Ahmet Sayar
Department of Computer Engineering, Kocaeli University
Umuttepe Campus, 41380, Kocaeli, Turkey
serifeacikalin9@gmail.com, {suleyman.eken, ahmet.sayar}kocaeli.edu.tr

## ABSTRACT

Processing and analysis of massive clinical data are resource intensive and time consuming with traditional analytic tools. Electroencephalogram (EEG) is one of the major technologies in detecting and diagnosing various brain disorders. Diagnosing sleep disorders are also possible by classifying the sleep stages from EEG signals. In this study, we investigate classification sleep stages using SparkMlib framewok and evaluate scalabilities of algorithms on physionet sleep records.

## 1. INTRODUCTION

Sleep disorder, or somnipathy, is a medical disorder of the sleep patterns of a person or animal. Some sleep disorders are serious enough to interfere with normal physical, mental, social and emotional functioning. Types of sleep disorders can be put into four groups: dysomnia, circadian rhythm sleep disorders, parasomnias, and medical or psychiatric disorder [1]. One technique for sleep disorder diagnosis is by analyzing the sleep stages. Using EEG and EMG (Electromyography), sleep stages will be visualized and analyzed.

EEG includes information about the work principles for both brain and body. The signals like EEG which is obtained by sensors. Obtained these signs can be interpreted directly by the doctors or they can be classified by programmers using some classification methods for diagnosis and treatment.

Sleep stages can be determined by the structural combination of electrical activities which produced by nerve cells. The sleep process includes different stages and these stages characterized by use of structural combination of electrical activities of nerve cells. Automatic approaches for sleep stage classification have been developed in many studies [2]-[6].

The aim of this work is to present sleep stage classification in distributed manner. In this study, sleep stages are classified into six different groups using SparkMLlib that is distributed machine learning framework. The success of developed system is determined by various metrics such as recall, precision and accuracy.

The rest of the paper is structured as follows. In the second section, preliminaries on sleep EEG and sleep stages, used dataset, and other sub-steps to process data, and classification algorithms are mentioned. Third section presents experimental evaluation. Conclusion and some future enhancements are given at the last section.

## 2. MATERIALS AND METHODS

### 2.1 Sleep EEG and Sleep Stages

Usually sleepers pass through five stages: 1, 2, 3, 4 and REM (rapid eye movement) sleep. These stages progress cyclically from 1 through REM then begin again with stage 1. A complete sleep cycle takes an average of 90 to 110 minutes. The first sleep cycles each night have relatively short REM sleeps and long periods of deep sleep but later in the night,

REM periods lengthen and deep sleep time decreases.

Stage 1 is light sleep where you drift in and out of sleep and can be awakened easily. In stage 2, eye movement stops and brain waves become slower with only an occasional burst of rapid brain waves. When a person enters stage 3, extremely slow brain waves called delta waves are interspersed with smaller, faster waves. In stage 4, the brain produces delta waves almost exclusively. Stages 3 and 4 are referred to as deep sleep or delta sleep and it is very difficult to wake someone from them. In the REM period, breathing becomes more rapid, irregular and shallow, eyes jerk rapidly and limb muscles are temporarily paralyzed. Table 1 shows frequency, amplitude and waveform type for each sleep stage.

Table 1. Characteristics of sleep stages

| Stage | Frequency (Hz) | Amplitude (µV) | Waveform type |
|---|---|---|---|
| awake | 15-50 | <50 | alpha rhthym |
| pre-sleep | 8-12 | 50 | beta rhthym |
| 1 | 4-8 | 50-100 | theta |
| 2 | 4-15 | 50-150 | splindle waves |
| 3 | 2-4 | 100-150 | spindle waves and slow waves |
| 4 | 0.5-2 | 100-200 | slow waves and delta waves |
| REM | 15-30 | <50 | alpha rhthym |

## 2.2 Dataset

The dataset for the experiment in this study was provided by MCH-Westeinde Hospital. The data can be downloaded from http://www.physionet.org/physiobank/database/sleep-edf/. This is a collection of 61 polysomnograms (PSGs) with accompanying hypnograms (expert annotations of sleep stages) from 42 subjects in two studies. The first was a study of age effects on sleep in healthy subjects (20 subjects, aged 25-34, with two 20-hour PSGs from consecutive nights for 19 subjects); the second was a study of temazepam effects on sleep in 22 subjects who had mild difficulty falling asleep but were otherwise healthy (9-hour PSGs of each subject on placebo) [7]-[7].

Dataset includes records of 79 healthy people between the ages of 25-101. Experts specify their sleep stages according to Rechtschaffen and Kales standard per 30 seconds.

## 2.3 Feature Extraction and Transformation of Sleep EEG Signals

Previous works about scoring and classification of sleep stages show that features of EEG signals are usually put in three groups: (i) time space features, (ii) frequency space features, and (iii) hybrid of time and frequency space features.

Feature extraction is done separately according to frequency range specified by Rechtschaffen and Kales. When all the data sets is considered, it contains about 500000000 (500 million) examples. Each sample includes 75 features and output class information/label is available for sample.15 features are extracted for 5 different frequencies. So, each sample has 75 (15x5) features. These features are: (i) arithmetic mean, (ii) harmonic mean, (iii) average value after outliers elimination, (iv) energy, (v) entropy, (vi-viii) minimum, median and maximum values, (ix) standard deviation, (x) skewness, (xi-xii) 0:25 quantile and 0.75 quantile, (xiii) inter quantile range, (xiv) skewness and (xv) kurtosis. Following sub-chapter explains multiclass classification and used algorithms.

## 2.4 Multiclass Classification

In machine learning, multiclass or polynomial classification is a problem that includes two or more class. Some classification algorithms allow using more than two class and some of them use binary algorithms. In this causes, using different strategies the conversion to polynomial classification is done.

### 2.4.1 Random Forest Algorithm (RF)

It is a sub-class of decision trees. It is one of the successful models of machine learning that used for method classification and regression. To reduce the risk of excessive compliance, multiple decision trees are used in algorithm [9]. Spark MLlib supports random forest algorithm for binary and multiclass classification and regression.

### 2.4.2 Gradient Random Forest Algorithm

It is a sub-class of decision trees. This method use decision trees in iterative training for reducing loss function. Spark MLlib sustains Gradient Random Forests algorithm binary classification and regression [9].

### 2.4.3 Adaptive Boosting (AdaBoost)

It is a meta-algorithm of machine learning and abbreviation is AdaBoost. It is used for the combination of learning algorithm to increase the performance of different type of learning algorithm. AdaBoost is sensitive to nosy and contradictory data [10].

### 2.4.4 Decision Trees (DT)

Decision trees that are used for binary and multiclass classification has continuous and categorical features. Leaf nodes are considered equivalent to the related K-class [11].

### 2.4.5 Naïve Bayes (NB)

The Bayesian classification represents a supervised learning method as well as a statistical method for classification. Assumes an underlying probabilistic model and it allows us to capture uncertainty about the model in a principled way by determining probabilities of the outcomes. It can solve diagnostic and predictive problems.

### 2.4.6 Support Vector Machine (SVM)

Support vector machines classification algorithm is one of the strongest and most successful one. It is based the idea to make the maximum limit. SVM supports both binary and multiclass classifications.

## 3. EVALUATION DESIGN AND EXPERIMENTS

It is investigated with metrics that classification process is how healthy. The basic concepts that used in assessing model performance are accuracy, precision, sensitivity and recall. The success of model is related to the number of examples that assigned to correct class and the number of examples that assigned to wrong class [11].

The information about these metrics is as following:

- Accuracy (A)

The most popular and simple way of measure the performance of model is accuracy rate. It is the ratio of the total number of examples that classified correctly and total number of examples.

$$\frac{TN+TP}{TN+TP+FN+FP} \qquad (1)$$

- Precision (P)

Precision is the ratio of the number of true positive examples to the total number of true positives and false positives.

$$\frac{TP}{TP+FP} \qquad (2)$$

- Recall (R)

It is the ratio of the number of positive examples that classified correctly and the total number of positive examples.

$$\frac{TP}{TP+FN} \qquad (3)$$

We test all aforementioned classification algorithms implemented in MLlib on sleep-edf dataset. Also, PCA (Principal Component Analysis) and SVD (Singular Value Decomposition) are applied to improve the performance of classification algorithms. Running times that observed from single node and multi-node are given in Tables 2-6.

Table 2. Performance Measure of Naive Bayes Classification Algorithm

| | Naive Bayes | A | P | R | Time (min) |
|---|---|---|---|---|---|
| On the Single Machine | C | 0.667 | 0.665 | 0.638 | 2.28 |
| | PCA | 0.216 | 1.0 | 0.216 | 2.01 |
| | SVD | 0.583 | 0.583 | 1.0 | 2.32 |
| On the More Than One Machine | S | 0.667 | 0.665 | 0.638 | 1.2 |
| | PCA | 0.216 | 1.0 | 0.216 | 1.0 |
| | SVD | 0.583 | 0.583 | 1.0 | 1.1 |

Table 3. Performance Measure of Logistic Regression Classification Algorithm

| | Logistic Regression | A | P | R | Time (min) |
|---|---|---|---|---|---|
| On the Single Machine | C | 0.823 | 0.730 | 0.886 | 3.49 |
| | PCA | 0.585 | 0.585 | 0.999 | 3.50 |
| | SVD | 0.821 | 0.905 | 0.967 | 4.02 |
| On the More Than One Machine | S | 0.823 | 0.730 | 0.886 | 3.1 |
| | PCA | 0.5851 | 0.585 | 0.999 | 1.7 |
| | SVD | 0.821 | 0.905 | 0.967 | 3.5 |

Table 4. Performance Measure of Decision Trees Classification Algorithm

| | DT | A | P | R | Time (min) |
|---|---|---|---|---|---|
| On the Single Machine | C | 0.819 | 0.760 | 0.825 | 2.32 |
| | PCA | 0.585 | 0.361 | 0.000 | 2.36 |
| | SVD | 0.725 | 0.668 | 0.658 | 3.35 |
| On the More Than One Machine | S | 0.819 | 0.760 | 0.825 | 1.2 |
| | PCA | 0.585 | 0.361 | 0.000 | 1.4 |
| | SVD | 0.725 | 0.668 | 0.658 | 1.4 |

Table 5. Performance Measure of Random Forest Algorithm

| | RF | A | P | R | Time (min) |
|---|---|---|---|---|---|
| On the Single Machine | C | 0.770 | 0.636 | 0.848 | 2.09 |
| | PCA | 0.584 | 0.584 | 1.0 | 3.05 |
| | SVD | 0.656 | 0.695 | 0.401 | 2.35 |
| On the More Than One Machine | S | 0.770 | 0.636 | 0.848 | 1.0 |
| | PCA | 0.5846 | 0.584 | 1.0 | 1.3 |
| | SVD | 0.656 | 0.695 | 0.401 | 1.3 |

Table 6. Performance Measure of Gradient Boosted Trees Classification Algorithm

| | GBT | A | P | R | Time (min) |
|---|---|---|---|---|---|
| On the Single Machine | C | 0.214 | 0.214 | 0.992 | 2.31 |
| | PCA | 0.215 | 0.215 | 1.0 | 2.35 |
| | SVD | 0.215 | 0.215 | 0.999 | 3.30 |
| On the More Than One Machine | S | 0.214 | 0.214 | 0.992 | 2.0 |
| | PCA | 0.215 | 0.215 | 1.0 | 2.01 |
| | SVD | 0.215 | 0.215 | 0.999 | 3.0 |

## 4. CONCLUSION AND FUTURE WORKS

Visual scoring for 8 hours sleeping takes lots of time and requires expertise work. Improvement of these automatic scoring system is designed for critical classification situations and helping doctors who do not have enough expertise. It saves time for doctors and minimizes the visual errors of experts.
In this study, automatic classification is done using distributed machine learning algorithms.

Socio-economic and psychological situations, age ranges, gender of sleep disorders may be elaborated for diagnosis and treatment of disease in the feature.

### Acknowledgments

This work has been supported by the TUBITAK under grant 1919B011503544.